\documentclass[12pt,preprint]{aastex}

\def\ms{m s$^{-1}$ }

\shorttitle{A super-Earth in the HZ of GJ 667C}
\shortauthors{Anglada-Escud\'e et al.}

\begin{document}

\title{A planetary system around the nearby M dwarf
GJ 667C with at least one super-Earth in its
habitable zone.}

\author{
Guillem Anglada-Escud\'e\altaffilmark{1,2}, %
Pamela Arriagada\altaffilmark{3},         %
Steven S. Vogt\altaffilmark{4},               %
Eugenio J. Rivera\altaffilmark{4},        %
R. Paul Butler\altaffilmark{1},           %
Jeffrey D. Crane\altaffilmark{5},            %
Stephen A. Shectman\altaffilmark{5},           %
Ian B. Thompson\altaffilmark{5},             %
Dante Minniti\altaffilmark{3,6,7},             %
Nader Haghighipour\altaffilmark{8}
Brad D. Carter\altaffilmark{9},              %
C. G. Tinney\altaffilmark{10},             %
Robert A. Wittenmyer\altaffilmark{10},        %
Jeremy A. Bailey\altaffilmark{10},            %
Simon J. O'Toole\altaffilmark{11},
Hugh R.A. Jones\altaffilmark{12},            %
James S. Jenkins\altaffilmark{13}             %
}

\email{anglada@dtm.ciw.edu}

\altaffiltext{1}{Carnegie Institution of Washington, Department of Terrestrial
Magnetism, 5241 Broad Branch Rd. NW, Washington D.C., 20015, USA}

\altaffiltext{2}{Universit\"{a}t G\"{o}ttingen, Institut f\"ur Astrophysik, Friedrich-Hund-Platz 1,37077 G\"{o}ttingen, Germany}

\altaffiltext{3}{Department of Astronomy, Pontificia Universidad Cat\'olica de Chile, Casilla 306, Santiago 22, Chile}

\altaffiltext{4}{UCO/Lick Observatory, University of California, Santa Cruz, CA 95064, USA}

\altaffiltext{5}{Carnegie Observatories, 813 Santa Barbara St., Pasadena, CA 91101-1292, USA}

\altaffiltext{6}{Vatican Observatory, V00120 Vatican City State, Italy}

\altaffiltext{7}{Department of Astrophysical Sciences, Princeton University, Princeton NJ 08544-1001}

\altaffiltext{8}{Institute for Astronomy \& NASA Astrobiology Institute,
University of Hawaii-Monoa, 2680 Woodlawn Drive, Honolulu, HI 96822, USA}

\altaffiltext{9}{Faculty of Sciences, University of Southern Queensland, Toowoomba, 4350, Australia}

\altaffiltext{10}{Department of Astrophysics, School of Physics, University of New South Wales, Sydney 2052, Australia}

\altaffiltext{11}{Australian Astronomical Observatory, PO Box 296, Epping, 1710, Australia}

\altaffiltext{12}{Centre for Astrophysics Research, University
of Hertfordshire, College Lane, Hatfield, Herts, AL10 9AB, UK}

\altaffiltext{13}{Departamento de Astronom\'ia, Universidad de Chile, Camino El Observatorio 1515, Las Condes, Santiago, Chile}

\begin{abstract}

We re-analyze 4 years of HARPS spectra of the
nearby M1.5 dwarf GJ 667C available through the
ESO public archive. The new radial velocity (RV)
measurements were obtained using a new data
analysis technique that derives the Doppler
measurement and other instrumental effects using
a least-squares approach. Combining these new 143
measurements with 41 additional RVs from the
Magellan/PFS and Keck/HIRES spectrometers,
reveals 3 additional signals beyond the
previously reported 7.2-day candidate, with
periods of 28 days, 75 days, and a secular trend
consistent with the presence of a gas giant
(Period$\sim$ 10 years). The 28-day signal
implies a planet candidate with a minimum mass of
4.5 M$_\earth$ orbiting well within the canonical
definition of the star's liquid water habitable
zone, this is, the region around the star at
which an Earth-like planet could sustain liquid
water on its surface. Still, the ultimate water
supporting capability of this candidate depends
on properties that are unknown such as its
albedo, atmospheric composition and interior
dynamics. The 75-day signal is less certain,
being significantly affected by aliasing
interactions among a potential 91-day signal, and
the likely rotation period of the star at 105
days detected in two activity indices. GJ 667C is
the common proper motion companion to the GJ
667AB binary, which is metal poor compared to the
Sun. The presence of a super-Earth in the
habitable zone of a metal poor M dwarf in a
triple star system, supports the evidence that
such worlds should be ubiquitous in the Galaxy.


\end{abstract}


\section{Introduction}

The Doppler detection of extrasolar planets is
achieved by measuring the periodic radial
velocity variations induced in a star by the
presence of orbiting low-mass companions. The
Doppler signature of a gas giant planet with
orbital parameters similar to Jupiter is about
$10$ \ms over a period of 11 years. By
comparison, Earth's reflex barycentric pull on
the Sun corresponds to 8 c\ms. The most precise
spectrographs can deliver long-term stability at
the level of 1--3
\ms\citep{mayor:2011,vogt:2010}. This precision
is sufficient to detect candidates of a few Earth
masses in tight orbits around low-mass stars (M
dwarfs). A key requirement in reaching such
accuracy is the extraordinary calibration of both
the wavelength scales of the spectrum and the
instrumental point-spread-function. The two most
successful methods used to date are (1) the
Iodine cell technique \citep{butler:1996} and (2)
the construction of stabilized spectrographs fed
by optical fibers \citep{baranne:1996}.

In the first of these two approaches, the star's
light is passed through a transparent cell
containing Iodine gas at low pressure. The
absorption spectrum of Iodine is imprinted on the
star's light, tracing precisely the same optical
path as encountered by the star light in traversing
the spectrometer.  The combined spectrum of the
star and Iodine is then modeled to obtain a
simultaneous point-spread-function, wavelength and
Doppler shift solution for the stellar spectrum
\citep{butler:1996}. The stabilized spectrograph
approach relies on the construction of a vacuum-sealed
spectrometer fed with optical fibers that
produces a fairly constant instrumental profile and
long-term wavelength stability. Each night,
stabilized spectrographs are calibrated in
wavelength by feeding the light from a wavelength
standard source (e.g., Th/Ar lamp) through the same
fiber as the science targets. This approach is
exquisitely implemented by HARPS installed on the
European Southern Observatory 3.6-m telescope at La
Silla Observatory \citep{harps:construction}. HARPS
is a cross dispersed echelle spectrograph 
with a resolving power of $\lambda/\delta\lambda$=110 000,
and over the years has demonstrated 1 \ms long-term stability
\citep{pepe:2011,mayor:2011}.

Even though HARPS is probably the most precise
astronomical spectrometer ever built, the
Cross-Correlation Function (CCF) data analysis
method that has been commonly used to analyze
this data, is suboptimal in the sense that it
does not exploit the full Doppler information in
the stellar spectrum \citep{pepe:2002}. For this
reason, instead of using the CCF RVs provided by
the ESO archive, we use a least-square
template-matching method to derive new RV
measurements. Thanks to the instrumental
stability and the excellent wavelength
calibration provided by the HARPS-ESO data
reduction software, the model required to match
each observation to a high signal-to-noise-ratio
(SNR) template only needs to include a Doppler
offset and a multiplicative polynomial to correct
for the flux variability across each echelle
order. The template is obtained by co-adding all
the spectra after a preliminary RV measurement is
obtained using the highest SNR observation. The
least-squares matching technique has been used on
HARPS data before. An example are the RV
measurements on GJ 1214 (V=14.57) used to derive
the mass of the transiting super-Earth reported
in \citet{charbonneau:2009}. The performance and
description of our software tool, HARPS-TERRA
(Template Enhanced Radial velocity Re-analysis
Application) on a representative sample of stars
can be found in \citet{anglada:2012}.

\section{Observations}


One of the M dwarfs with many public HARPS
spectra but no published detections is GJ 667C.
The ESO archive contains 143 observations of this
star obtained between June 2004 and October 2008.
Typical exposure time is between 900 and 1500
seconds and the average signal-to-noise ratio
(SNR) is 64 at 6100 \AA. At a 2009 conference, a
planet candidate orbiting this star was announced
with a period $P$ of $\sim$7 days. Also recently,
\citet{bonfils:2011} reported the detection of a
plausible signal with $P\sim28$ days similar to
one of the candidates we report here. However, no
detailed analysis nor any data were provided
therein. We use our new HARPS-TERRA software 
to derive new RV measurements on GJ 667C
\citep{anglada:2012}. The root-mean-square (RMS)
of these RVs is 3.89 \ms, which is significantly
larger than the median internal precision ($\sim$
1.1 m $s^{-1}$) and the typical RMS found on
other stable M dwarfs \citep{bonfils:2011}. We
found that the CCF RV measurements provided by
the ESO archive were noisier (RMS $\sim$ 4.3 \ms)
causing  the signals reported in Section
\ref{sec:analysis} to appear less significant. In
order to obtain more secure detections, we
obtained 21 new measurements with the Planet
Finder Spectrograph (PFS) between August 2011 and
October 2011. PFS is a cross-dispersed echelle
spectrograph recently installed at the 6.5m Clay
Magellan Telescope at Las Campanas Observatory,
and uses the Iodine cell technique to obtain RV
measurements at 1--2 \ms precision
\citep{crane:2010}.

GJ 667C has also been observed with HIRES/Keck
\citep{vogt:1994} using the Iodine cell method for
just over a decade. Such long-time of observation
should deliver tighter constraints on long period
signals. However, in August 2004, the HIRES CCD array
was replaced and the data obtained prior to this
upgrade is of markedly inferior quality. Post-fix
HIRES measurements show similar scatter to the HARPS
and PFS RVs, so only 20 post-fix HIRES observations
were used in our analysis. We 
emphasize that the signals discussed here were first
detected using HARPS-TERRA measurements only, and
that the contribution of the PFS and HIRES data were
to improve the sampling cadence and increase the
significance of the detections. All the RV
measurements used in our analysis are given in the
on--line version of the manuscript.

\section{Properties of GJ 667C}


According to \citet{skiff:2010}, GJ 667C (HR 6426 C)
is classified as an M1.5 dwarf. The star is a common
proper motion companion to the K3V+K5V binary GJ
667AB \citep[HR 6426AB, ][]{kron:1957}. At the
distance to the system ($\sim$ 6.8 pc), the minimum
physical separation between GJ 667C and GJ667 AB is
$\sim$230 AU. The metallicity of GJ 667AB has been
measured before \citep[e.g.][]{perrin:1988} and
amounts to [Fe/H] = -0.59$\pm$0.10, meaning that the
system is metal poor compared to the Sun. The same
studies show that the GJ 667AB pair is well within
the main sequence, indicating an age between 2 and 10
Gyr \citep[e.g.][]{strobel:1981}. The membership of
GJ 667 to a Galactic population is unclear. Although
its low metallicity points to a thick disk
membership, its total velocity in the Galactic Local
Standard of Rest is rather low (44.6$\pm$1.5 k\ms),
which is characteristic of thin disk kinematics
\citep[see Figure 3 in][]{bensby:2003}. The most
recent parameters of GJ 667AB can be found in
\citet{cvetkovic:2011} and \citet{tokovinin:2008}.
Using the empirical relations given by
\citet{delfosse:2000}, the HIPPARCOS parallax of the
GJ 667AB pair \citep{hipparcos:2007} and K band
photometry from 2MASS \citep{twomass:2006}, we
derived a mass of 0.310$\pm$0.019 M$_{\sun}$ for GJ
667C. The luminosity and effective temperature of GJ 667C are
derived from the models in \citet{baraffe:1998} by
assuming the aforementioned mass, metallicity and an
age of 5 Gyr.

\section{Orbital analysis} \label{sec:analysis}

Keplerian orbital fits to the combined RV data were
obtained using the SYSTEMIC interface
\citep{systemic}, which allows the interactive
least-squares adjustment of complex multiplanetary
systems to several data sets. To determine whether
there was a significant periodicity remaining in the
data, we used a custom-made version of a
least-squares periodogram \citep{cumming:2004} that
adjusts a separate zero-point offset to each
instrument (HARPS,PFS, and HIRES).

\begin{figure}[tb]
\centering
\includegraphics[width=4.5in,clip]{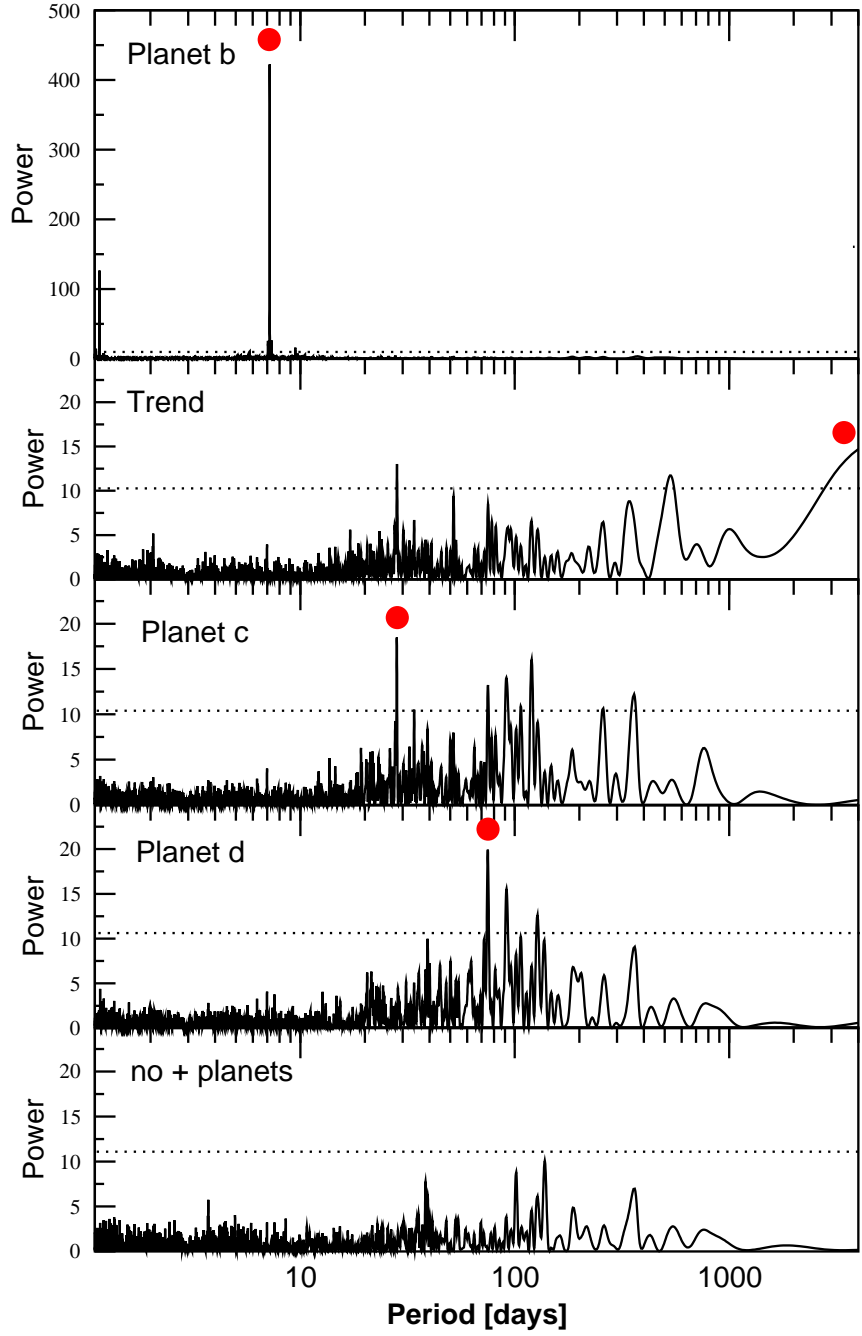}

\caption{Detection periodograms of the three candidate
planets and long period trend detected in the RV
measurements of GJ 667C. The signals are listed from
top to bottom in order of
detection.}\label{fig:periodograms}

\end{figure}

To quantify the significance of a new signal, we
estimated its False Alarm Probability (FAP)
empirically. We created 10$^5$ synthetic sets by
randomly permutating the RV measurements over the
same observing epochs (while retaining membership
within each instrument). We then computed the
periodogram of each synthetic set. A
\textit{false alarm} was identified when a
synthetic data set produced a periodogram peak
higher than the power of the signal under
inspection. The number of false alarms was then
divided by the number of simulations to derive
the FAP, which was used as a measure of the
probability that a spurious detection arised due
to an unfortunate arrangement of the noise. This
method is widely used to assess the likelihood of
periodic signals and a detailed description can
be found elsewhere \citep[e.g.,][]{cumming:2004}.
Since a few tens of M dwarfs have been
intensively followed up at 1 \ms precision
\citep{bonfils:2011} and to minimize the chances
of detecting a false positive, only signals with
a FAP $<1\%$ were added to the solution  (dotted
lines in Figure\,\ref{fig:periodograms}).

The first detected signal was an extremely
significant periodicity at 7.2\,days (see Figure
\ref{fig:periodograms}). No false alarms were found
in any of the 10$^5$ synthetic sets indicating a
FAP $<$ 0.001\%. The signal corresponds to a planet
with a minimum mass ($M \sin i$) of 5.2 M$_\earth$
(GJ 667Cb) and a slightly eccentric orbit.

\begin{figure}[tb]
\centering
\includegraphics[width=5.0in,clip]{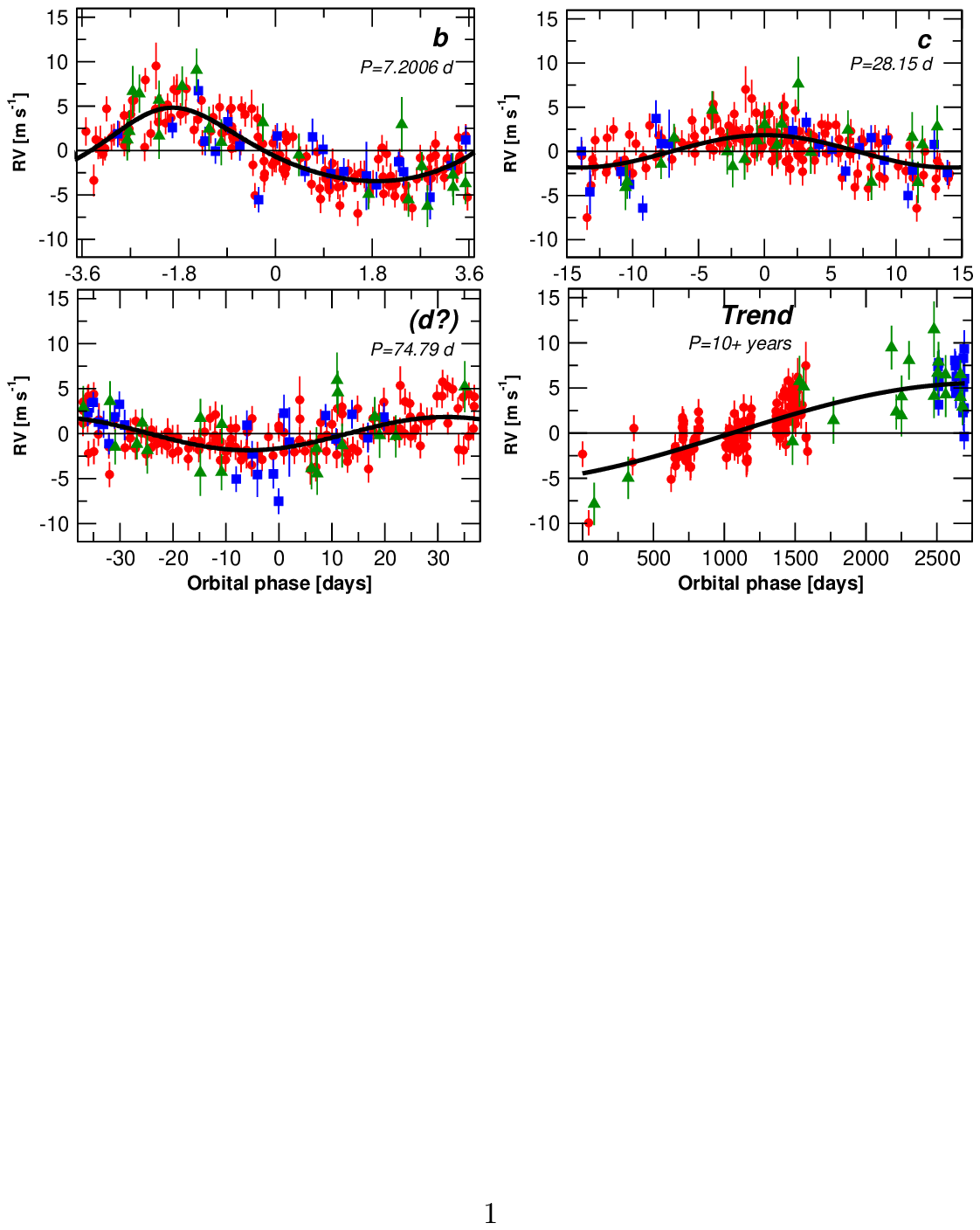}

\caption{Phase-folded RV measurements of the four
signals discussed in the text. The 143 HARPS
measurements are shown in red circles, 21 PFS measurements
are shown in blue squares and the green triangles correspond to
the 20 HIRES observations. Each preferred Keplerian
model is shown as a black line.}
\label{fig:phase_folded}

\end{figure}

After subtracting a Keplerian solution for GJ
667Cb, a secular trend was the next most
significant signal. The magnitude of the trend
($\sim$ 1.8 \ms yr$^{-1}$) is compatible with the
gravitational pull from the GJ 667AB pair
(maximum value is $\sim$ 3.6 \ms yr$^{-1}$) but
could also be caused by an additional unseen
long-period companion. The corresponding FAP of
this signal is 0.055\%, a very significant
detection. A tentative solution with a period of
7000 days provides a slight improvement to the
fit due to some curvature detected when combining
HIRES+PFS measurements (see Figure
\ref{fig:phase_folded}). We estimate that one
more year of observations is required to
determine if the signal is due to an additional
low mass companion or due to the gravitational
pull of the GJ 667AB binary.

\begin{figure}[tb]
\centering
\includegraphics[width=6.0in,clip]{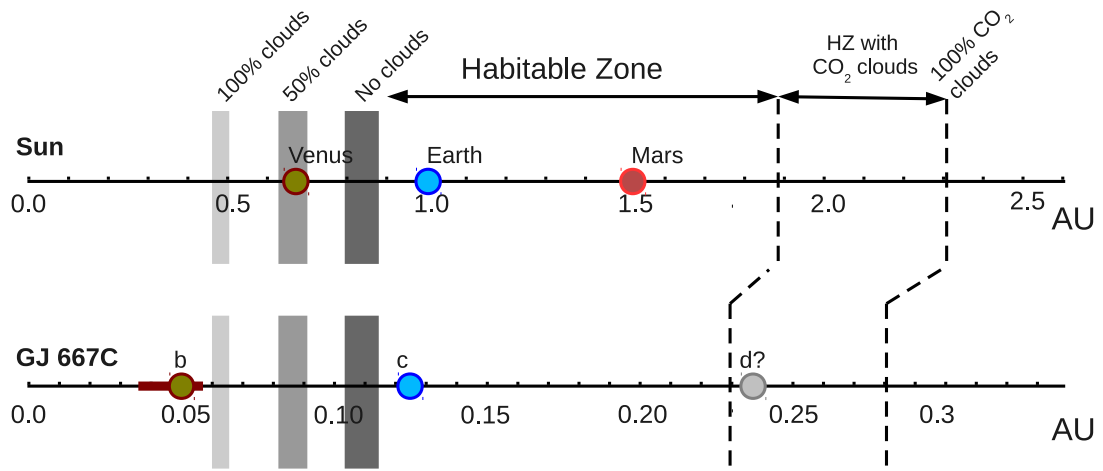}

\caption{Comparative liquid water habitable zones
for the Sun and GJ 667C \citep{selsis:2007}. The 
gray areas indicate
the theoretical inner edge for different
fractional cloud coverage. The outer edge is
marked with a dashed line. The actual
habitability of GJ 667Cc depends on physical
parameters that are currently unknown.}

\label{fig:topview}

\end{figure}

The next signal also has a very low FAP
(0.034\%), and implies a planet with a period of
28.15 days and $M\sin{i}\sim4.5\,M_{\earth}$ (GJ
677Cc). Although the period is close to the lunar
aliasing frequency ($\sim$ 27.3 days), the
orbital phase coverage is complete thanks to the
multi-year time-span of the observations. Because
of its small amplitude, eccentric solutions
cannot be ruled out \citep{shen:2008,
otoole:2009}. A Monte Carlo Markov Chain analysis
\citep{ford:2005} indicates that this
eccentricity (with 98\% confidence) must be less
than 0.27. At a semi-major axis of 0.123 AU, the
stellar flux $S$ reaching the top of its
atmosphere is $90\%$ of the solar flux received
by Earth (S$_0$). Using $L$ and $T_{eff}$ for the
host star, the relations given in
\citet{kane:2011} provide boundaries of the zone
at which liquid water could exist on an
Earth-like planet (also called liquid water
habitable zone, or HZ). In the canonical model
presented in \citet{kasting:1993} and updated in
\citet{selsis:2007}, the inner and outer
boundaries depend on the fractional cloud
coverage for the putative planet and are
displayed as thick gray rectangles in Figure
\ref{fig:topview}. Even though these limits are
uncertain, GJ 667Cc comfortably falls in this HZ
and also satisfies the empirical limits set by an
unhabitable Venus and a possibly habitable early
Mars \citep{selsis:2007}. Let us remark that the
ultimate capability of GJ 667Cc to support liquid
water depends on properties that are not yet
known (e.g, albedo, atmospheric composition and
interior dynamics). Detailed studies using 
realistic climatic and geodynamical models
\citep[e.g.,][]{heng:2011,wordsworth:2011} are
needed to better assess its chances of supporting
life.

After fitting for GJ 667Cc, we found a group of
candidate periodicities between 75 and 105 days.
This time domain is strongly affected by aliases.
For example, HARPS data alone favors a period of
91 days, which is uncomfortably close to a very
clear signal detected in two activity indices
(Section \ref{sec:activity}). When combining all
the data, the 75-day periodicity had the lowest
FAP (0.021\%), indicating that such signal could
not be ignored in the analysis. A tentative
orbital solution  assuming a circular orbit is
given in Table \ref{tab:solution}. A putative
super-Earth with P$\sim$75 days could also
support liquid water if its atmosphere contained
high concentrations of CO$_2$ \citep[e.g., GJ
581d in][]{wordsworth:2011}. We reiterate that,
until more observations become available, this
signal should be considered with due caution.

In a more thorough analysis (not presented here for
brevity), we examined other orbital solutions
with up to 4 signals at alternative periods
(including, among others, signals at periods of
105, 120, and 33\,days). All these attempts
delivered significantly poorer fits, extreme
eccentricities, and planetary systems that were
unstable on time scales shorter than 1 Myr.

We also performed long-term N-body simulations
based on some of our fits using the Hybrid
simplectic integrator included in the Mercury
integration package \citep{chambers:1999}. We
included the first order partial Post-Newtonian
correction in the central star's gravitational
potential as in \citet{lissauer:2001} and used a
time step of 0.2 days. We assumed a coplanar
system throughout. Since the ratio of the periods
of the inner two planets is near 4:1, we checked
the four critical angles (involving only the
periastron longitudes, $\omega$) for the 4:1 mean
motion resonance (MMR). We also examined the
difference in the periastron longitudes. We found
that the critical angles for the MMR circulate,
whereas the difference in the longitudes of
periastron librates about $180 \deg$ with
amplitude $\sim 90\deg$. Results of our
simulations also show that the eccentricities of
the planets librate between 0.0 and 0.235 for
companion b and between 0.04 and 0.265 for
companion c in opposite senses as a result of
angular momentum conservation. Thus, the system
appears to be protected by a secular resonance
between the two inner planets in which 1) the
orbits can become nearly anti-aligned when the
eccentricity of b is small, and 2) the periastron
longitudes are nearly perpendicular when the
eccentricity of b is large.  Most of the time,
the system is in the second configuration (in
good agreement with the fitted parameters in
Table 1). This stabilizing mechanism appears to
function for at least the first 25 Myr of our
simulations. Further research into the dynamical
evolution of this system is warranted once the
nature of the other two signals (long-period
trend and the 75/91-day candidate) is better
understood.

\section{Periodic signals in the activity indicators}
\label{sec:activity}

Here we show the analysis of the time series of
three activity indicators : the Bisector span
(BIS), the full-width-half-maximum of the CCF
(FWHM) and the CaII H+K S-index in the Mount
Wilson system (S-index). Measurements of the BIS
and the FWHM were provided by the HARPS-ESO
pipeline. The S-index  \citep{baliunas:1995} was
directly measured on the blaze-corrected spectra
using the definitions given by
\citet{lovis:2011}. Since the BIS and FWHM could
not be obtained in the Iodine cell approach, we
limited our analysis to the HARPS observations
only. Briefly, the BIS is a measure of the
asymmetry of the average spectral line and should
correlate with the RV if the observed offsets are
caused by spots or plages rotating with the star
\citep{queloz:2001}. The FWHM is a measure of the
width of the mean spectral line and its
variability is usually associated with changes in
the convective patterns on the surface of a star.
\citet{lovis:2011} found that the FWHM is the
most informative index when investigating the
correlation of stellar activity with RV
variability on cool stars (T$_{eff}<4600$ K). The
S-index is an indirect measurement of the
chromospheric emission which depends upon the
intensity of the stellar magnetic field. Because
the strength of the magnetic field affects the
efficiency of convection, the S-index could also
correlate with observed RV variability. Since the
connection between activity and RV jitter on M
dwarfs is not well understood at the few \ms
level \citep{lovis:2011}, we limit our analysis
to evaluate if any of the indices has a
periodicity that could be related to any of the
detected candidates.

While the BIS did not show any significant
periodicity, the S-index and especially the FWHM
did show significant power around 105 days. To
obtain meaningful periodograms, one outlier point
had to be removed in the FWHM (a 9.4-$\sigma$
outlier at JD=2454677.66) and another one from
the S-index (a 13-$\sigma$ outlier at
JD=2454234.79). The FAPs of such signals were
obtained by applying same empirical method used
for the RVs. We found that the 105-day peak in
the S-index had a FAP of 0.2\% and no false
alarms were found in any of the synthetic sets
generated for the FWHM (FAP$<$0.001\%). Given the
low metallicity and the age estimates of the GJ
667AB binary, GJ 667C should have a slow rotation
rate \citep{irwin:2011} compatible with the
105-day signal observed in these two indices.
Because of the significant aliases affecting the
$\sim$ 100-day time domain, it is likely (but not
conclusive) that GJ 667Cd is a spurious signal
resulting from localized magnetic activity
rotating with the stellar surface. In addition to
more RV observations to improve the phase
sampling, a photometric follow-up could help to reveal 
the true nature of GJ 667Cd.

\section{Conclusions}

We have derived precision RV measurements from
public HARPS spectra using a least-squares
matching approach on the M dwarf GJ 667C and
thereby detected the Doppler signature of (at
least) two planets. Additional observations with
PFS and HIRES confirm the detection of these
signals and further constrain the orbital
parameters. Even though the public CCF Doppler
measurements are not as precise, the CCF method
still provides useful information on stellar
activity that can be used to investigate the
origin of candidate signals. 

GJ 667Cc is the super-Earth candidate most
securely detected within the liquid water
habitable zone of another star. As for other
proposed candidates \citep[e.g., GJ 581d and HD
85512b announced in][respectively]{mayor:2009,pepe:2011}, 
its actual capability of supporting liquid
water depends on many physical properties that are yet
unknown.  Using the relations given by
\citet{charbonneau:2007}, the reported
candidates have non-negligible probabilities
of transiting in front of the star ($\sim
2.7\%, 1.1\%$, and $0.6\%$ for planets b, c,
and d, respectively). Even though these
probabilities are low, the estimated transit
depth assuming a density similar to Earth, is
about 0.3\%, which can be measured using small
aperture telescopes. Statistical
extrapolations based on Doppler, transit and
microlensing surveys indicate that such
planets should be abundant around main
sequence stars
\citep{mayor:2009,borucki:2011,cassan:2012}.
With the new generation of optical and
infrared spectrographs, many nearby M dwarfs
will be efficiently surveyed for low mass
planets. If the detection rate holds, very
soon now we may have a real chance of
searching for spectroscopic signatures of
water and life on one of these worlds.

\textbf{Acknowledgements} We thank
the constructive comments given by the 
anonymous referee. This research has been
partially supported by the Carnegie Postdoctoral
Program to GAE; ICM P07-021-F, FONDAP  15010003
and BASAL-CATA PFB-06 grants to DM and PA; NSF
grant AST-0307493 to SSV; NASA NNX07AR40G and
NASA Keck PI program grants to RPB; NASA grants
NNA04CC08A and NNX09AN05G to NH; ARC grant
DP774000 to CGT; and Fondecyt grant 3110004 to
JSJ. This work is based on data obtained from the
ESO Science Archive Facility. Observations were
obtained from Las Campanas Observatory and W. M.
Keck Observatory. W. M. Keck Observatory is
operated jointly by Univ. of California and
California Institute of Technology. This research
has made use of the SIMBAD database, operated at
CDS, Strasbourg, France.

\begin{figure}[tb]
\centering
\includegraphics[width=6.0in,clip]{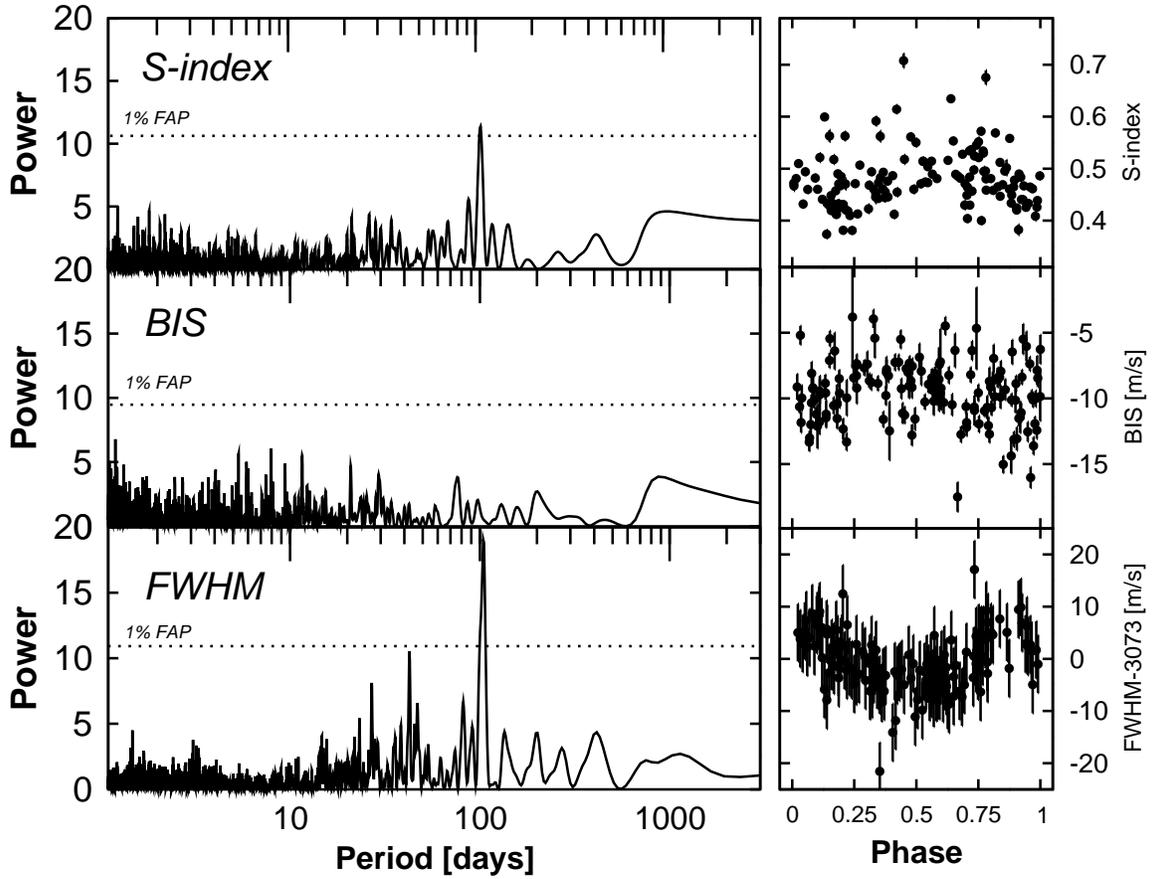}

\caption{Left: Periodograms of the three activity
indicators discussed in the text. Both the S-index
and the FWHM show a significant signal around 105
days. On the right, we show each activity indicator
folded to the most significant period : 105
days for the S-index and the FWHM, and 1.2008 days
for the BIS.} \label{fig:activity}

\end{figure}

\begin{deluxetable}{lrrrr}  
\tabletypesize{\scriptsize}

\tablecaption{Best Keplerian solution to the
planetary system around GJ 667C. The numbers in
parenthesis indicate the uncertainty in the last two
significant digits of each parameter value.
Uncertainties have been obtained using a Bayesian
MCMC analysis \citep{ford:2005} and represent the
68\% confidence levels around the preferred solution.
All orbital elements are referred to JD$_0$ =
2453158.7643. The properties of the star are listed
at the bottom (references given in the text).}

\tablehead{
  \colhead{Parameter} &
  \colhead{b} &
  \colhead{c} &
  \colhead{(d?)} &
  \colhead{Trend}
}

\startdata
P [days]                       & 7.20066(67)    & 28.155(17)   &  74.79(12)   & 7100(3000)   \\ 		
$M \sin i$ [$M_{jup}$]         & 0.01789(75)    & 0.0143(12)   & 0.0178(17)   & 0.25(12)     \\ 		
$M \sin i$ [$M_{\Earth}$]      &    5.68(23)    &   4.54(38)   &   5.65(54)   & 79(40)	     \\			
$M_0$ [deg]                    &  106.6(3.5)    &    144(25)   &    211(11)   & 231(10)      \\ 		
e                              &   0.172(43)    &    $<$0.27   &   0(fixed)   & 0(fixed)     \\ 		
$\omega$[deg]                  &     344(12)    &    238(20)   &   0(fixed)   & 0(fixed)     \\ 		
Detection FAP                  &   $<0.001\%$   &   $0.034\%$  &  $ 0.021\%$  & $0.055\%$    \\ 		
K [\ms]                        & 3.90           & 2.02	       & 1.84	      & 8.41     \\			
a [AU]                         & 0.049          & 0.123(20)    & 0.235        & 2.577    \\			
S/S$0$                         & 570\%          & 90.5(3.0)\%  & 24.8\%       & --    \\			
                                                                                             \\
Statistics                                                                                   \\
\hline                                                                     	             \\		
$N_{\rm HARPS}$     & 143 & &Total $N_{\rm obs}$		      &     184    \\
RMS$_{HARPS}$ [\ms] & 1.89& &RMS [\ms]			              &    2.05    \\
$N_{\rm PFS}$	    & 21  & & $ \chi^2$			              &  310.99    \\
RMS$_{PFS}$ [\ms]   & 2.37& & $\chi^2_{\nu}$                          &    1.88    \\
$N_{\rm HIRES}$     & 20  &                                           &            \\
RMS$_{HIRES}$ [\ms] & 2.85&                                           &            \\
                                                                                   \\
Star parameters              &               & &Derived quantities                  \\
\hline                                                                             \\
R.A.                         &  17 18 57.16  & &Mass [M$_\sun$]   & 0.310(19)         \\
Dec                          & -34 59 23.14  & &Spectral type     & M1.5V             \\
$\mu_{R.A.}$ [mas yr$^{-1}$] &  1129.7(9.7)  & &UVW$_{\rm LSR}$ [k\ms] & (19.5, 29.4,-27.2) \\
$\mu_{Dec.}$ [mas yr$^{-1}$] &  -77.02(4.6)  & &Age estimate      & $>2$ Gyr          \\
Parallax [mas]               &  146.29(9.0)  & &$T_{eff}$[K]      & 3700$\pm$100      \\
Hel. RV [k\ms]               &     6.5(1.0)  & &$L_*$/L$_\sun$    & 0.01370(90)       \\
${\rm[Fe/H]}$                &    -0.59(10)  & &$R^{HZ}_{in}$[AU] & 0.1145(72)        \\
V [mag]                      &    10.22(10)  & &$R^{HZ}_{out}$[AU]& 0.226(14)         \\
K [mag]                      &    6.036(20)  &                                       \\
\enddata
\label{tab:solution}
\end{deluxetable}


\begin{thebibliography}{43}
\expandafter\ifx\csname natexlab\endcsname\relax\def\natexlab#1{#1}\fi

\bibitem[{{Anglada-Escud{\'e}} \& {Butler}(submitted)}]{anglada:2012}
{Anglada-Escud{\'e}}, G., \& {Butler}, R.~P. submitted to ApJS, --,

\bibitem[{{Baliunas} {et~al.}(1995){Baliunas}, {Donahue}, {Soon}, \&
  et~al.}]{baliunas:1995}
{Baliunas}, S.~L., {Donahue}, R.~A., {Soon}, W.~H.,  {et~al.} 1995, \apj, 438,
  269

\bibitem[{{Baraffe} {et~al.}(1998){Baraffe}, {Chabrier}, {Allard}, \&
  {Hauschildt}}]{baraffe:1998}
{Baraffe}, I., {Chabrier}, G., {Allard}, F., \& {Hauschildt}, P.~H. 1998, \aap,
  337, 403

\bibitem[{{Baranne} {et~al.}(1996){Baranne}, {Queloz}, {Mayor}, {Adrianzyk},
  {Knispel}, {Kohler}, {Lacroix}, {Meunier}, {Rimbaud}, \&
  {Vin}}]{baranne:1996}
{Baranne}, A., {et~al.} 1996, \aaps, 119, 373

\bibitem[{{Bensby} {et~al.}(2003){Bensby}, {Feltzing}, \&
  {Lundstr{\"o}m}}]{bensby:2003}
{Bensby}, T., {Feltzing}, S., \& {Lundstr{\"o}m}, I. 2003, \aap, 410, 527

\bibitem[{{Bonfils} {et~al.}(2011){Bonfils}, {Delfosse}, {Udry}, \&
  et~al.}]{bonfils:2011}
{Bonfils}, X., {Delfosse}, X., {Udry},  {et~al.} 2011, e-prints arXiv:1111.5019

\bibitem[{{Borucki} {et~al.}(2011){Borucki}, {Koch}, {Basri}, \&
  et~al.}]{borucki:2011}
{Borucki}, W.~J., {Koch}, D.~G., {Basri}, G.,  {et~al.} 2011, \apj, 736, 19

\bibitem[{{Butler} {et~al.}(1996){Butler}, {Marcy}, {Williams}, \&
  et~al.}]{butler:1996}
{Butler}, R.~P., {Marcy}, G.~W., {Williams}, E.,  {et~al.} 1996, \pasp, 108,
  500

\bibitem[{{Cassan} {et~al.}(2012){Cassan}, {Kubas}, {Beaulieu}, \&
  et~al.}]{cassan:2012}
{Cassan}, A., {Kubas}, D., {Beaulieu}, J.-P.,  {et~al.} 2012, \nat, 481, 167

\bibitem[{{Cayrel de Strobel}(1981)}]{strobel:1981}
{Cayrel de Strobel}, G. 1981, Bulletin d'Information du Centre de Donnees
  Stellaires, 20, 28

\bibitem[{{Chambers}(1999)}]{chambers:1999}
{Chambers}, J.~E. 1999, \mnras, 304, 793

\bibitem[{{Charbonneau} {et~al.}(2009){Charbonneau}, {Berta}, {Irwin}, \&
  et~al.}]{charbonneau:2009}
{Charbonneau}, D., {Berta}, Z.~K., {Irwin}, J.,  {et~al.} 2009, \nat, 462, 891

\bibitem[{{Charbonneau} {et~al.}(2007){Charbonneau}, {Brown}, {Burrows}, \&
  {Laughlin}}]{charbonneau:2007}
{Charbonneau}, D., {Brown}, T.~M., {Burrows}, A., \& {Laughlin}, G. 2007,
  Protostars and Planets V, 701

\bibitem[{{Crane} {et~al.}(2010){Crane}, {Shectman}, {Butler}, \&
  et~al.}]{crane:2010}
{Crane}, J.~D., {Shectman}, S.~A., {Butler}, R.~P.,  {et~al.} 2010, in SPIE
  Conference Series, Vol. 7735

\bibitem[{{Cumming}(2004)}]{cumming:2004}
{Cumming}, A. 2004, \mnras, 354, 1165

\bibitem[{{Cvetkovic} \& {Ninkovic}(2011)}]{cvetkovic:2011}
{Cvetkovic}, Z., \& {Ninkovic}, S. 2011, VizieR On-line Data Catalog:
  J/other/Ser/180.71, 4201, 18001

\bibitem[{{Delfosse} {et~al.}(2000){Delfosse}, {Forveille}, {S{\'e}gransan}, \&
  et~al.}]{delfosse:2000}
{Delfosse}, X., {Forveille}, T., {S{\'e}gransan}, D.,  {et~al.} 2000, \aap,
  364, 217

\bibitem[{{Ford}(2005)}]{ford:2005}
{Ford}, E.~B. 2005, \aj, 129, 1706

\bibitem[{{Heng} \& {Vogt}(2011)}]{heng:2011}
{Heng}, K., \& {Vogt}, S.~S. 2011, \mnras, 415, 2145

\bibitem[{{Irwin} {et~al.}(2011){Irwin}, {Berta}, {Burke}, \&
  et~al.}]{irwin:2011}
{Irwin}, J., {Berta}, Z.~K., {Burke}, C.~J.,  {et~al.} 2011, \apj, 727, 56

\bibitem[{{Kane} \& {Gelino}(2011)}]{kane:2011}
{Kane}, S.~R., \& {Gelino}, D.~M. 2011, \apj, 741, 52

\bibitem[{{Kasting} {et~al.}(1993){Kasting}, {Whitmire}, \&
  {Reynolds}}]{kasting:1993}
{Kasting}, J.~F., {Whitmire}, D.~P., \& {Reynolds}, R.~T. 1993, \icarus, 101,
  108

\bibitem[{{Kron} {et~al.}(1957){Kron}, {Gascoigne}, \& {White}}]{kron:1957}
{Kron}, G.~E., {Gascoigne}, S.~C.~B., \& {White}, H.~S. 1957, \aj, 62, 205

\bibitem[{{Lissauer} \& {Rivera}(2001)}]{lissauer:2001}
{Lissauer}, J.~J., \& {Rivera}, E.~J. 2001, \apj, 554, 1141

\bibitem[{{Lovis} {et~al.}(2011){Lovis}, {Dumusque}, {Santos}, {Bouchy},
  {Mayor}, {Pepe}, {Queloz}, {S{\'e}gransan}, \& {Udry}}]{lovis:2011}
{Lovis}, C., {et~al.} 2011, ArXiv e-prints

\bibitem[{{Mayor} {et~al.}(2009){Mayor}, {Bonfils}, {Forveille}, \&
  et~al.}]{mayor:2009}
{Mayor}, M., {Bonfils}, X., {Forveille}, T.,  {et~al.} 2009, \aap, 507, 487

\bibitem[{{Mayor} {et~al.}(2011){Mayor}, {Marmier}, {Lovis}, \&
  et~al.}]{mayor:2011}
{Mayor}, M., {Marmier}, M., {Lovis}, C.,  {et~al.} 2011, ArXiv e-prints

\bibitem[{{Meschiari} {et~al.}(2009){Meschiari}, {Wolf}, {Rivera}, \&
  et~al.}]{systemic}
{Meschiari}, S., {Wolf}, A.~S., {Rivera}, E.,  {et~al.} 2009, \pasp, 121, 1016

\bibitem[{{O'Toole} {et~al.}(2009){O'Toole}, {Tinney}, {Jones}, \&
  et~al.}]{otoole:2009}
{O'Toole}, S.~J., {Tinney}, C.~G., {Jones}, H.~R.~A.,  {et~al.} 2009, \mnras,
  392, 641

\bibitem[{{Pepe} {et~al.}(2002){Pepe}, {Mayor}, {Galland}, \&
  et~al.}]{pepe:2002}
{Pepe}, F., {Mayor}, M., {Galland},  {et~al.} 2002, \aap, 388, 632

\bibitem[{{Pepe} {et~al.}(2003){Pepe}, {Rupprecht}, {Avila}, \&
  et~al.}]{harps:construction}
{Pepe}, F., {Rupprecht}, G., {Avila}, G.,  {et~al.} 2003, in SPIE Conference
  Series, Vol. 4841, 1045--1056

\bibitem[{{Pepe} {et~al.}(2011){Pepe}, {Lovis}, {S{\'e}gransan}, {Benz},
  {Bouchy}, {Dumusque}, {Mayor}, {Queloz}, {Santos}, \& {Udry}}]{pepe:2011}
{Pepe}, F., {et~al.} 2011, \aap, 534, A58+

\bibitem[{{Perrin} {et~al.}(1988){Perrin}, {Cayrel de Strobel}, \&
  {Dennefeld}}]{perrin:1988}
{Perrin}, M.-N., {Cayrel de Strobel}, G., \& {Dennefeld}, M. 1988, \aap, 191,
  237

\bibitem[{{Queloz} {et~al.}(2001){Queloz}, {Henry}, {Sivan}, {Baliunas},
  {Beuzit}, {Donahue}, {Mayor}, {Naef}, {Perrier}, \& {Udry}}]{queloz:2001}
{Queloz}, D., {et~al.} 2001, \aap, 379, 279

\bibitem[{{Selsis} {et~al.}(2007){Selsis}, {Kasting}, {Levrard}, \&
  et~al.}]{selsis:2007}
{Selsis}, F., {Kasting}, J.~F., {Levrard}, B.,  {et~al.} 2007, \aap, 476, 1373

\bibitem[{{Shen} \& {Turner}(2008)}]{shen:2008}
{Shen}, Y., \& {Turner}, E.~L. 2008, \apj, 685, 553

\bibitem[{{Skiff}(2010)}]{skiff:2010}
{Skiff}, B.~A. 2010, VizieR Online Data Catalog, 10, 2023

\bibitem[{{Skrutskie} {et~al.}(2006){Skrutskie}, {Cutri}, {Stiening},
  {Weinberg}, {Schneider}, {Carpenter}, {Beichman}, {Capps}, {Chester},
  {Elias}, {Huchra}, {Liebert}, {Lonsdale}, {Monet}, {Price}, {Seitzer},
  {Jarrett}, {Kirkpatrick}, {Gizis}, {Howard}, {Evans}, {Fowler}, {Fullmer},
  {Hurt}, {Light}, {Kopan}, {Marsh}, {McCallon}, {Tam}, {Van Dyk}, \&
  {Wheelock}}]{twomass:2006}
{Skrutskie}, M.~F., {et~al.} 2006, \aj, 131, 1163

\bibitem[{{Tokovinin}(2008)}]{tokovinin:2008}
{Tokovinin}, A. 2008, \mnras, 389, 925

\bibitem[{{van Leeuwen}(2007)}]{hipparcos:2007}
{van Leeuwen}, F. 2007, \aap, 474, 653

\bibitem[{{Vogt} {et~al.}(1994){Vogt}, {Allen}, {Bigelow}, \&
  et~al.}]{vogt:1994}
{Vogt}, S.~S., {Allen}, S.~L., {Bigelow}, B.~C.,  {et~al.} 1994, in SPIE
  Conference Series, ed. {D.~L.~Crawford \& E.~R.~Craine}, Vol. 2198, 362

\bibitem[{{Vogt} {et~al.}(2010){Vogt}, {Butler}, {Rivera}, \&
  et~al.}]{vogt:2010}
{Vogt}, S.~S., {Butler}, R.~P., {Rivera}, E.~J.,  {et~al.} 2010, \apj, 723, 954

\bibitem[{{Wordsworth} {et~al.}(2011){Wordsworth}, {Forget}, {Selsis}, \&
  et~al}]{wordsworth:2011}
{Wordsworth}, R.~D., {Forget}, F., {Selsis}, F.,  {et~al.} 2011, \apjl, 733,
  L48

\end{thebibliography}
\end{document}